# An awareness-based model to minimize the environmental damage of the internet usage: A Longitudinal Study

*Completed Research Paper*

**Ayodhya Wathuge**
Southern Cross University
Queensland, Australia
ayodhyawathuge@gmail.com

**Darshana Sedera**
Southern Cross University
Queensland, Australia
darshana.sedera@gmail.com

## Abstract

*The record-breaking increase of internet usage in 2020 with the spread of the COVID-19 pandemic has made us think about the alarming consequences of it in the aspect of climate change. As countries go into lockdown the use of the internet to perform tasks remotely has increased in record numbers. As per the trend and at times addictive nature of its usage, it is unlikely that the usage of the internet will decrease in the future reducing its current contribution to climate change. Considering the sustainability perspective, this study investigates whether the pervasive nature of internet usage could be reduced by simply inducing awareness. A population-based survey experiment comprising of 326 respondents was employed to investigate if awareness alone could reduce individual internet usage.*

## Introduction

The year 2020 showed a staggering growth of internet usage especially with the COVID-19 pandemic (Håkansson 2020; Sun et al. 2020). With the restrictions in mobility and lock downing of countries and cities around the world, people turned towards alternative methods to perform daily tasks. Therein, demand for remote working, learning, and other services increased rapidly as the countries struggle to stabilize economies from collapsing (NHS 2020). The work trend is now bending towards performing tasks remotely, and it will escalate with the pandemic. A study conducted in France, Germany, Spain, UK, and the US shows that average online content consumption doubled from 3 hours and 17 minutes to nearly 7 hours in 2020 with the pandemic (DoubleVerify 2020). Although there are signs of the COVID-19 pandemic ending with a global vaccination program, there are clear signs of internet use keep moving up. In a live-streamed meeting, Facebook's Chief Executive Officer, Mark Zuckerberg has stated that he expects remote working to be growing in a post-COVID era (Conger 2020). Moreover, the internet use has increased as the majority of individuals acknowledge the internet as a medium that makes lives simpler and the environment sustainable. Presently, reading from screens, and video conferencing over traveling have attracted more attention (Moberg et al 2010). Therein, usage of the internet has escalated as the internet provides countless benefits to individuals as well as to the environment.

Although the purported benefits of internet use are boundless, it also is a major contributor to climate change. Researchers have estimated that one Google search emits approximately 20 milligrams of carbon dioxide ($CO_2$) (Hölzle 2009). This figure is compelling, considering the daily number of Google Searches and YouTube views performed, which are approximately one and two billion respectively (Gombiner 2011). Berners-Lee estimates the annual $CO_2$ emission of the internet to 300 million tons which is equal to a year's oil, gas, and coal burnings of Turkey or Poland (Berners-Lee 2011). Moreover, an email is estimated to add at least four grams of $CO_2$ (Berners-Lee 2011). Further, it is estimated that 6.8 grams of carbon are generated each time a web site is loaded, which equals to $CO_2$ emitted in boiling water for a cup of tea (Clifford 2019). Therein, simple behavioral changes like downloading vs. streaming, frequency of checking of social media, unnecessary and spam emails beg the question of the impact of individual internet practices.

Although there are overwhelming statistics to demonstrate the environmental pollution caused by the internet (Lokuge et al. In Press; Sedera et al. 2017b), *public awareness* about the matter is scant. Ironically, the majority of individuals use the internet to search on reducing environmental pollution not knowing that the act itself damages the environment. According to Google Trends, the global average percentage of searches of the phrase "internet pollution" is 2% for the last five years (2015-2020) while the phrase "environmental pollution" scores 69% showing that the public has more awareness towards environmental pollution and specifically, a very less knowledge on internet pollution. Moreover, as per the articles referred related to internet pollution, the majority of the papers have addressed on making devices and processes energy efficient (Briscar 2017; Molla 2009; Salahuddin et al. 2016), acquiring renewable energy to power data centers (Goiri et al. 2013), and performing different green IT practices in organizational contexts (Molla 2009; Nishant et al. 2012; Nishant et al. 2013; Tushi et al. 2014), while some studies have simply acknowledged about the existence of the internet pollution (Park et al. 2018). Studies conducted to create awareness of internet pollution are very few (El Idrissi and Corbett 2016). El Idrissi and Corbett (2016) in their literature analysis show that awareness is still in the phase of maturation. Moreover, Jenkin et al. (2011) show that even organizations are in the initial stages of awareness of green information technology practices, implying that awareness of environmental pollution of the internet is very low. Therefore, creating an awareness of the environmental damage that we do, knowingly or unknowingly, could be an important perspective of damaging climate change effects.

In defining awareness, consciousness plays an important part. Consciousness is the general capacity that an individual possesses for a certain kind of subjective experience. Awareness is the demonstration of this general capacity (Tulving 1993). Once awareness is instilled in someone, individuals tend to modify their behaviors based on their level of awareness (Kang et al. 2012). Past studies demonstrate the effectiveness of heightening awareness to modify behavior. Among them, the literature on anti-smoking, road safety, drink driving, and speed driving show a successful integration of awareness to alter behaviors. For example, a trend of ceasing smoking has been observed with the smokers who are open to quitting, upon the awareness provided through mass media awareness campaigns (Vallone et al. 2011). Further, awareness campaigns designed to suit different age groups and gender have shown significant results in altering driver attitudes and behaviors (Gauld et al. 2020). However, there are counter-arguments that some behaviors are difficult to change simply by creating awareness. For example, prior studies acknowledge that some behaviors, especially corruptive habits or addictive behaviors are harder to change simply with awareness (Carey et al. 2013). Accordingly, does the great dependence and at times addictive nature of the internet allows an individual to change their behaviors simply because of creating environmental awareness?

In answering the above question, this article investigates the possibility of awareness to reduce internet usage in order to minimize carbon emissions. First, the study investigates theoretical and practical implications related to awareness and behavior change and proposes a model to address the study objective. Secondly, the methodology and subsequently, an empirical analysis is conducted using data from an experiment that demonstrates the before and after effects of awareness. Next, the results are presented. Finally, the article concludes with the conclusion, limitations, and future implications.

## Theoretical and Practical Implications: Awareness and Consumer Behavior

Given its importance in multi-disciplines, there are a plethora of mature theories that have looked at how awareness leads to behavior changes. They explain the process of behavior change in various ways. We have considered a few contexts that show similarity to internet usage, which are pervasive and hard to abandon. They are smoking, drink driving, road safety, and speed driving. In selecting studies, we looked at a representative sample of articles that are based on a theory. (Cohen et al. 2007; Delaney et al. 2004) have identified that theory of reasoned action (TRA), the theory of planned behavior (TPB), social cognitive theory (SCT), self-determination theory (SDT), and elaboration likelihood model (ELM) are among the popularly used theories to depict awareness leading to action[1].

Due to page limitations we only explain two of the most widely used theories in this regard: TPB and SDT. TPB was developed based on TRA, one of the oldest theories that investigate the relationship between awareness and behavior change (Hale et al. 2002). TPB extends the scope of TRA to facilitate explaining behavior where individuals feel less control over whether the volition occurs or not. According to TPB, behaviors take place due to behavioral intention that occurs due to three precursors: attitudes, subjective norm and perceived behavioral control (Ajzen 1991). On the other hand, in SDT long-lasting behaviors are expected to take place when the goals and values of a person are more internalized (Ryan and Deci 2000). The intrinsic motivation provided in the form of awareness, education or feedback that enhance relatedness, competence and/or autonomy are among primary methods that increase internalization (Anderson and Rodin 1989; Koo and Chung 2014). Moreover, the referred studies have defined awareness in multiple ways. Chatzisarantis and Hagger in their study based on TPB define awareness as "the extent to which people are aware of the stimulus triggering a process or of the process itself" (Chatzisarantis and Hagger 2007). Further, Bulgurcu et al. define awareness as the "overall knowledge and understanding of potential issues" (Bulgurcu et al. 2010). Extending the same concept some studies have referred to awareness as the extent to which a message 'stimulates cognitive activity' (Kreuter and Wray 2003). More commonly awareness is referred to as consciousness or knowledge about an event or the ability to recall things (Goel et al. 2011).

Awareness about an event or an issue can be of different types. We have identified that past studies have prominently used one or more types of the following three types of awareness in modifying behaviors. They are problem or consequence awareness, awareness of remedial measures, tools or techniques, and awareness of government, organization, or community-specific rules and regulations. Problem or consequence awareness provided in contexts such as speed driving, drink driving and smoking through mass-media like radios, newspapers, and packaging have generated effective results in mitigating such behaviors (Frieden et al. 2005; Rundmo and Iversen 2004). A study carried out in Malaysia about eco-labeled product introduction has shown that the public has reacted positively to embrace environmentally less harmful products like biodegradable cleaning agents and recycled paper showcasing type two awareness (Rashid 2009). Moreover, Chen et al. in their study show that awareness created by signboards of upcoming speed cameras has increased the relationship between penalty strategy and speed driving (Chen et al. 2020). Furthermore, in the context of water conservation, water usage bans have been shown to achieve increased positive results depicting the awareness of government rules (Barrett 2004).

When inducing awareness, past studies have observed the type of awareness as well as the outcomes of it. The approach of Kruger and Kearney, Gundu and Flowerday (2013) explains that awareness creates knowledge and it causes attitudes to change consequently influencing the behavioral intention, behavior, and finally the culture. (Stern 2000; Stern et al. 1995) explain the same process in a more elaborative way. They posit that the knowledge gained through awareness generates or intensifies personal values, leading to the generation of a specific set of beliefs, attitudes, and intention to change behavior (Sedera and Dey 2013). Further, Han and Ryu (2009) proclaim that changing behavior and adhering to it would lead to loyalty, where an individual becomes devoted in performing a specific behavior. Overall, these studies intend to demonstrate the change to one's behavior as a result of heightened awareness.

---

[1] Besides, theories and models such as the Extended parallel process model, Transtheoretical model, Deterrence theory, Terror management theory and Protection motivation theory also provide useful insights concerning the discussion.

The studies have also investigated the moderators of awareness led behavior. Among them, self-efficacy, gender, level of education, maturity based on age in creating awareness occupy a prominent place (Alshboul and Streff 2017; Preston et al. 2000; Ziadat 2010). Moreover, the type of message (Abroms et al. 2012), the wording of message (Lewis et al. 2008), quality, and fairness of awareness campaigns also have been demonstrated to have moderating effects on behavior change (Alshboul and Streff 2017; Sedera and Lokuge 2018; Sedera et al. 2017a).

Table 1 illustrates the theories used in quitting smoking, drink driving prevention, road safety, and speed driving contexts with few example studies that show the relationship between awareness and behavior change. The contexts of the studies as well as the relationships observed show similarity to the current study. Context wise, they show resemblance to internet usage as they are addictive and corruptive as at times the internet usage although they are considered as bad habits. On the other hand, the studies have focused on a common independent and dependent variable which are awareness and behavior change respectively. Moreover, those contexts have incorporated strategies more than awareness such as pricing strategies (taxes and fines) and legal measures to minimize such behaviors. It allows the researcher to investigate the problem in hand with different perspectives, unlike comparing with other contexts such as internet usage (where pricing or legal measures have not been popularly used) and sustainability (not a pervasive behavior) where strategies more than awareness have been rarely used.

The studies in Table 1 imply that awareness and internet usage are two important variables that complement each other. Therefore, we argue that to minimize rapidly increasing data demand it is important to study the relationship between awareness of internet environmental pollution and internet usage to determine whether awareness alone can create an impact.

## The Conceptual Model

Humans are centrally concerned with motivation— in doing a task or making another person do a task. The conceptual framework of the research is developed reflecting on the SDT. According to SDT, long-lasting behaviors are expected to take place when the goals and values of a person are more internalized (Ryan and Deci 2000). There are two basic ways in motivating an individual namely, extrinsic and intrinsic motivation. SDT suggests that a person may engage in behavior due to external reasons such as receiving rewards and punishments or/and internal reasons such as the satisfactory feeling of performing an action (Darner 2009; Pelletier 2002).

Past studies in the contexts of quit smoking, drink and speed driving have used SDT in explaining how awareness changes behavior. They have used various mechanisms to create awareness. Choi et al (2014) in their study suggest that information addressing reasons for smoking cessation, step-wise guides, and alert or alarm messages that increase the autonomy could guide smokers towards quitting the practice. Moreover, in a longitudinal study (Pesis-Katz et al. 2011) show that information about smoking cessation, treatment programs and discussions about health led smokers attend tobacco abstinence. (Choi and Noh 2019) show that smoking cessation campaigns affected smokers' perception of stigma and motivation for cessation. Furthermore, (Williams et al. 2009; Williams et al. 2011) also support the same argument.

The SDT is used successfully to achieve road safety behaviors. (Ranjit et al. 2017) in their study conducted in Nepal show that directive messages influenced individuals to understand the value in road safety behaviors which in turn affected in increased behavioral intention. Moreover, (Watson-Brown et al. 2020) investigate learner teaching approaches to minimize road accidents. (Watson-Brown et al. 2019) further claim that higher-order driving instructions delivered to develop a self-regulated safety orientation, have potential to reduce young novice drivers' risky driving behaviors. Using SDT, these studies show how awareness created through different mechanisms can change and minimize addictive behaviors. The key tenet of SDT is that individuals experience motivation to do an activity once their three basic psychological needs are satisfied. In the study, a video is used to satisfy the three basic psychological needs of a person. In satisfying autonomy need which is to create knowledge about causes and effects and to enable people to take better decisions, we provided information on environmental damage of the internet. For example, the video provided information about the current overall $CO_2$ emission of the internet and its estimated impacts to help people understand the effects of using internet. In satisfying the need for competence which is to be cognitively and affectively related to the activities the individuals do, we provided information related to the activities they do on the internet and their $CO_2$ percentage, which instill competence in them to reduce such activities. Thirdly, in satisfying the relatedness need which is to perceive that an individual is related and is accepted by the society, we included statements that deliver the message, reducing internet usage and environmental damage is important and is an accepted virtue in the society.

According to the theory, environmental awareness provides knowledge arousing personal values, providing a sense of choice and volition internalizing the motivation (Ryan and Deci 2008). We define awareness as the end stage experience that results from the filtering and processing of the possible set of experiences (Vaneechoutte 2000). Following that we define environmental awareness as the independent variable and internet usage as the dependent variable. Here we assess the constructs of SDT implicitly. Further, we assess the potential impact of age and gender, given that the past literature shown to be moderating the effects of age (Ng et al. 2012) and gender (Dohnke et al. 2011) on the relationship between awareness led behavior (however, not in the current context). Figure 1 shows the conceptual model of the study.

The study is designed as an experiment. The majority of studies in Table 1 have used the same method to investigate the changes to human behavior. Therefore, following the same strategy, this study also employs a population-based survey experiment to investigate the relationship between awareness and internet usage.

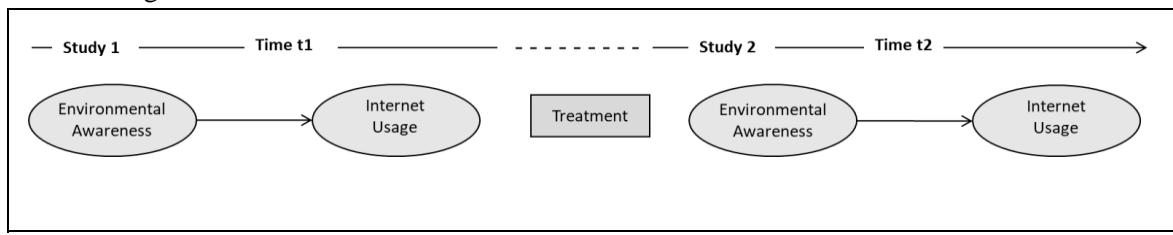

**Figure 1. The Study Design**

## Experiment Design

To test the model, a population-based survey experiment consisting of questions related to awareness and internet usage was employed. The survey consisted of questions related to demographic characteristics of respondents, awareness of internet damage on the environment: measured using the perception they hold on environmental damage, and internet usage: measured using time spent on the internet, the willingness to change current usage pattern, frequency of use, and addiction to it. A seven-point Likert scale was used to measure the variables where 1 represented Extremely Disagree and 7 represented Extremely Agree.

| Table 1. Table of Sample Studies | | | | | |
|---|---|---|---|---|---|
| **Context** | **Dependent Variable** | **Independent Variable** | **Theory** | **Studies** | **Description** |
| Quit Smoking | After the information campaigns about quitting smoking, the studies measured the current smoking status of individuals as well as the beliefs, attitudes, and behaviors towards quit smoking in longitudinal studies conducted at different points of time. | The studies manipulated the variables such as pro-tobacco receptivity, tobacco independence, awareness of health risks, awareness of industry targeting, and social consequences through different types of information campaigns. | TRA | (Hershey et al. 2005; Syu et al. 2010) | Among the studies referred, SCT has been used by most researchers to determine the relationship between awareness and behavior change |
| | The studies measured the variables such as motivation and intention to quit, substance and general assertiveness in quit smoking generally through longitudinal surveys conducted before and after intervention programs | Studies have investigated the effect intervention programs did on attitudes, subjective norms, and perceived behavioral control that resulted to modify the dependent variable. | TPB | (Dohnke et al. 2011; Godin et al. 1992; Zhao et al. 2019) | |
| | Among the dependent variables motivation to quit, readiness, and confidence to quit were popularly measured in different time points after the intervention program. | Variables such as the stage of change, self-efficacy to quit, beliefs about smoking, engagement with the program, and acceptability of the program were measured upon creating awareness. | SCT | (Dino et al. 2004), (Woodruff et al. 2007),(Whittaker et al. 2011), (Abroms et al. 2012), (Bock et al. 2013),(Maisto et al. 1999) | |
| | Among the dependent variables motivation to quit smoking was | Variables such as campaign exposure, type of | SDT | (Pesis-Katz et al. 2011), (Choi and Noh 2019), | |
| | tested after carrying out different intervention programs | intervention were measured as independent variables | | (Williams et al. 2009; Williams et al. 2011) | |
| Drink driving | Attitudes and intentions relating to drink driving were assessed at two-time points after advertisement viewing. | The effectiveness and usefulness of advertisements were assessed to test the response efficacy of advertisements | ELM | (Lewis et al. 2008) | Most of the studies have used TPB as a theoretical model to determine attitude and behavioral changes |
| | Intention to drink and drive was tested at different points in time before and after the intervention program was carried out | Attitudes and normative beliefs towards drinking were measured before and after the intervention program to assess the attitude change. | TPB | (Nathanail and Adamos 2013), (Adamos and Nathanail 2016; Sheehan et al. 1996) | |
| Road safety and speed driving | The studies have tested the intention to drive within speed limits, behavioral changes in following road safety measures, and intention to reduce speed driving at different points in time after the intervention programs were carried out. | The studies have manipulated the variables such as pre-driver beliefs (education), attitudes, subjective norms, and perceived behavioral control towards speed driving through different information campaigns. | TPB | (Poulter and McKenna 2010), (Steinmetz et al. 2016), (Cuenen et al. 2016), (McKenna 2007) | Studies explaining theoretical aspects of awareness and behavior change are comparatively less |
| | As dependent variable the studies have measured intention to drive safely. | Type of message and teaching approaches are manipulated to test the effect they create on the knowledge of drivers. | SDT | (Ranjit et al. 2017), (Watson-Brown et al. 2020), (Watson-Brown et al. 2019) | |

We adapted instruments, by reviewing literature related to the constructs and identifying different ways that they can be conceptualized to best capture the aspects of awareness, and internet usage. Questions related to measuring awareness were adapted from studies such as (Bulgurcu et al. 2010; Morgil et al. 2004). The question items were as follows: My internet usage can be damaging to the environment, I am aware of the cost that data centers have on the environment, I would like my internet providers and services to use renewable energy sources and changing my internet usage habits can help reduce environmental problems. Questions about internet usage were derived from (Mathwick and Rigdon 2004; Venkatesh et al. 2008). The question items were as follows: On average I spend many hours per week on the internet, compared with my friends, I spend more time on the internet, outside the time I spend with emails, I consider myself a heavy internet user, I frequently have multiple devices connected to my internet and I try limit my time using the internet, but found it challenging. The Study-1 consisted of questions about the current awareness of internet damage and current internet usage.

After the Study-1 survey data collection was completed, we induced the participants with a tailor-made video of four minutes about the damage that internet consumption does to the environment. Through keen observation of several videos, we selected a video published by a public news channel of France named France 24 on YouTube, as it contained relevant content proved in the literature that could create awareness, and also the video length was appropriate. The video was embedded in the questionnaire itself. It provided information about the amounts of carbon emissions of most popular internet-based activities such as circulating emails and web searches. Moreover, it introduces the concept of data centers and their energy requirements. Overall, the video provides a succinct awareness of the damage internet does to the environment. To capture the number of people who have watched the video before and after the questionnaire was given, we kept a count on the number of YouTube views.

Then Study-2 survey was executed after the video to capture the cognitive responses evoked. It included questions to assess the attitudinal change that occurred after watching the video. There, the current intention to use the internet was questioned. Therein, we test the role of the targeted awareness campaigns (Maibach 1993), which is purported to be altering behaviors.

### The Sample

To test the model in Figure 1, a representative sample was drawn from a population of ordinary internet users. It was necessary that the sample comes from a single geographical area, have a similar educational background, and have similar economic conditions to increase the homogeneity and to control the effect of extraneous variables. As such, a stratified sample of 326 internet users based on gender were recruited by randomly distributing the questionnaire link among social media users of Sri Lanka. The global internet user gap of male vs female is 17% as per the International Telecommunication Union (ITU 2019). According to our sample, the gap between internet using male (58.3%) and female (41.7%) was 16.6% which approximately equals the global internet user gender gap. Respondents of ages between 22-57 responded to the survey. All the respondents could read and understand English. A larger sample would be more appealing to the generalization of the results. However, the current sample does well represent the global internet users who use the internet on average 6 hours and 42 minutes a day which approximately resembles a Sri Lankan's average internet usage time (Kemp 2019). The majority of the respondents in the sample were young adults. Therefore, the sample represented a trustworthy number of major internet users, because global internet usage is more skewed towards young adults and their general awareness is less. We selected Sri Lanka as the context to conduct the study as it falls among the top ten countries that offer the lowest data charges (Howdle 2020).

## Results

For the analysis, we employed two types of tests: partial least squares (PLS) structural equation modeling (SEM) method (Benitez et al. 2017) and multi-group analysis (MGA) to understand the effects of age and gender groups. PLS was used as it allows to better understand the increasing complexity by exploring theoretical extensions that we aim to make related to the SDT (Hair et al. 2019). Moreover, the causal predictive approach in PLS-SEM helps provide the causal explanations of variables used in the study (Benitez et al. 2017). In the latest update of PLS method, Benitez proclaims that studies with small samples (less than 90 subjects) might produce less accurate statistics (Benitez et al. 2017). However, as the current study consists of 326 subjects, the mentioned drawback of the method is overlooked. First, we conducted a confirmatory composite analysis to test the overall fit of the proposed research models (Figure 1) (Benitez et al. 2017)[2]. It followed the guidelines of (Dijkstra 2010) and sub-constructs were estimated by using the regression weights (mode B). Therein, we observed the standardized root mean squared residual (SRMR), unweighted least squares (ULS) discrepancy (dULS), and geodesic discrepancy (dG), which let us assess the appropriateness of the composite model. The SRMR of the a-priori model was 0.026 for Study-1 and 0.031 for Study-2 – well below the recommended threshold of less than 0.080 – at the 0.05 alpha level (Benitez et al. 2017).

Next, we established construct validity using the average variance extracted (AVE), where we observed that for each construct the AVE is greater than the variance shared between the construct and the other constructs in the model, thus indicating strong construct and discriminant validity.

Finally, we observed strong and significant path coefficients from awareness to internet use (as shown in Figure 1. The path coefficients between awareness and use for Study-1 was 0.412 and for Study-2 it was 0.655. The $R^2$ for internet use Study-1 was 0.29 and Study-2 revealed an $R^2$ of 0.39. The extent that awareness is highly correlated with internet usage means that the conditions of SDT (i.e., autonomy, competence and relatedness) are satisfied implying that people are motivated to reduce internet usage. Next, we conducted a multi-group analysis to test whether age and gender (the pre-defined subpopulations) generate significant heterogeneous observations (Henseler 2007). Therein, we observed no significant differences between those that are under 25 years and those who are over ($| (\beta^{Age = <25} - \beta^{Age > 25}) | = 0.073$; t-value = 0.88; p-value = 0.437). However, in relation to gender, we observed significant differences between male vs. female in relation to the effects of environmental awareness on internet use ($| (\beta^{male} - \beta^{female}) | = 0.114$; t-value = 5.193; p-value = 0.001).

## Conclusion

This study examined the likelihood of using awareness as the only strategy to minimize internet usage. Based on prior literature, internet usage is referred to as an addictive behavior that is pervasive and hard to give up. Using the responses of 326 internet users the study concludes that using awareness as the only strategy to reduce internet usage shows statistically significant results. While the study findings look tautological from the outset, it reveals several important findings: (i) the environment awareness can lead to reduced internet use, that will eventually reduce climate damage, (ii) environmental awareness can be increased with simple communications, (iii) the higher environmental awareness leading to lower internet use is non-sensitive to the respondent age, but (iv) sensitive to one's gender.

The results are consistent with the literature to a certain extent. First, the study results show that awareness modifies behavior alone. The reason would be that the internet damage is comparatively a new concept than other already known pervasive behavioral contexts which incorporate combined strategies in controlling behavior such as awareness and price regulations. However, long-term impact of using awareness alone to minimize internet usage needs to be investigated as internet usage is projected to be drastically increasing. Contingent upon the past studies we suggest investigating awareness coupled with other regulations such as pricing interventions or government rules related to sustainable consumption, to examine whether a combined effect provides more influential results. Secondly, internet usage can be reduced by simple communication strategies such as creating awareness of the damage, other than using intensified awareness strategies such as fear generating messages or in-depth conversations which have been employed in other pervasive behavioral contexts. Thirdly, the higher environmental awareness leading to lower internet use is non-sensitive to the respondent's age. It may be because youth show an increasing attention towards sustainability of the environment which equalizes the environmental concerns of adults. However, gender shows sensitivity towards the observed relationship. Males are more sensitive may be because of their comparatively high engagement in the community and political issues than their female counterparts.

---

[2] Both models were tested using ADANCO 2.0.1 software with the bootstrap re sampling method (4,999 re samples).

Moreover, our results conform to the SDT theoretical underpinnings. According to the SDT, awareness as a regulation works well towards the high end of the continuum where an individual has the power of choice and volition. The results of the study show that awareness of environmental damage of internet has increased self-determination of internet users to minimize the usage despite any external regulation such as price changes or legal regulations.

The study has several limitations. First, this research was tested using data only from Sri Lanka, which would impact on generalizability, restricting of gaining an overall global picture of the changes in internet usage in relation to awareness. Therefore, we encourage further validation of results by gathering data from other countries that have various internet infrastructure and cultural values, which may affect the apprehension of awareness. Moreover, the data were collected in 2020 during a period of COVID-19 pandemic, where the usage of the internet could be relatively higher than normal, as internet was conceived as an indispensable asset substituting mobility. Therefore, the perspective on internet usage could have been simply exaggerated. As such, future research can be conducted to identify if the internet usage was exaggerated at the time of conducting the research. Additionally, other intervention strategies such as price and tax increases can be tested with awareness to measure the difference between behavior changes that occur due to awareness alone and awareness used with price increases.